# Effects of inorganic seed promoters on MoS$_2$ few-layers grown via chemical vapor deposition

Alessandro Cataldo[1,2], Pinaka Pani Tummala[1,3,4], Christian Martella[1], Carlo Spartaco Casari[5], Alessandro Molle[1], Alessio Lamperti[1,+]

[1] *CNR-IMM, Agrate Brianza Unit, via C. Olivetti 2, Agrate Brianza, I-20864, Italy*

[2] *Dipartimento di Chimica, Materiali e Ingegneria Chimica, Politecnico di Milano, P.zza Leonardo da Vinci 32, edificio 6, I-20133 Milano, Italy*

[3] *Department of Physics and Astronomy, University of Leuven, Celestijnenlaan 200D, B-3001 Leuven, Belgium*

[4] *Interdisciplinary Laboratories for Advanced Materials Physics (I-LAMP), Dipartimento di Matematica e Fisica, Università Cattolica del Sacro Cuore, via della Garzetta 48, 25133 Brescia, Italy*

[5] *Dipartimento di Energia, Politecnico di Milano; via Ponzio 34/3, I-20133 Milano, Italy*

+ Corresponding Author: alessio.lamperti@cnr.it

*Abstract*

In the last years, transition metal dichalcogenides (TMDs), especially at the two-dimensional (2D) limit, gained a large interest due to their unique optical and electronic properties. Among them, MoS$_2$ received great attention from the scientific community due to its versatility, workability, and applicability in a large number of fields such as electronics, optoelectronics and electrocatalysis. To open the possibility of 2D-MoS$_2$ exploitation, its synthesis over large macroscopic areas using cost-effective methods is fundamental. In this study, we report a method for the synthesis of large-area (~ cm$^2$) few-layers MoS$_2$ via liquid precursor CVD (L-CVD), where the Mo precursor (i.e. ammonium heptamolybdate AHM) is provided via a solution that is spin-coated over the substrate. Given the capability of organic and inorganic molecules, such as alkaline salts, to enhance MoS$_2$ growth, we investigated the action of different inorganic salts as seed promoters. In particular, by using visible Raman spectroscopy, we focused on the effect of Na(OH), KCl, KI, and Li(OH) on the thickness, morphology, uniformity and degree of coverage of the grown MoS$_2$. We optimized the process tuning parameters such as the volume of spin-coated solution, the growth temperature, and the seed promoter concentration, to synthesise the lowest possible thickness which resulted to be 2 layers (2L) of the highest quality. We witnessed that the addition of an inorganic seed promoter in the solution improves the extension of the grown $MoS_2$ promoting lateral growth front, and therefore the degree of coverage. From this study, we conclude that, amongst the investigated seed promoters, K-based salts proved to grant the growth of high-quality two-layer $MoS_2$ with optimal and uniform coverage of the $SiO_2$/Si substrate surface.

## 1. INTRODUCTION

Since the discovery of graphene in 2004, the interest towards 2D materials beyond graphene has rapidly increased through the years. Amongst them, Transition Metal Dichalcogenides (TMDs), at the 2D scale, have been, and still are, the subjects of increasing research interest [1], [2]. The focus on $MoS_2$ arose from several reasons: its easy synthesis and handling and, more importantly from a physical and engineering perspective, its tuneable bandgap [3]–[6]. Nonetheless, one of the major issues to address before industrial 2D-$MoS_2$ exploitation is the control of its growth over large areas, targeting the wafer scale. For this reason, in the last decade, many efforts have been directed towards engineering 2D-$MoS_2$ synthesis to obtain a high quality and uniform few-to-mono layer $MoS_2$, targeting a facile and scalable process [7], [8].

Chemical Vapour Deposition (CVD) is a versatile and efficient technique to synthesize TMDs, particularly $MoS_2$, down to ultrathin thicknesses [9]. In a typical CVD-based process for the synthesis of $MoS_2$, Mo precursor and S are introduced in the form of powders. Once introduced in the reactor, they are vaporized, and chemical reactions occur in the gas phase leading to the deposition of $MoS_2$ on a chosen substrate. This process is highly effective in obtaining $MoS_2$ monolayer triangular domains of several $\mu m^2$ areas but presents some limitations. First, the lack of control of resulting thickness and uniformity over large areas ($cm^2$) and, second, a relatively poor substrate surface coverage. To overcome these limitations, organic molecules, such as perylene-3,4,9,10-tetracarboxylic acid tetra-potassium salt (PTAS) have been exploited to condition the substrate surface before the growth to act as promoters of enhanced 2D growth of $MoS_2$ during the CVD process [4], [10]–[13].

Keeping in mind these concepts, the main target of this work is to synthesize and characterize few-layers $MoS_2$ over large macroscopic areas via liquid precursor atmospheric pressure CVD (L-CVD) and, in particular, to focus on the effect of inorganic seed promoters in the growth process. L-CVD involves the delivery of the Mo precursor and seed promoters starting from the formulation of a solution that is deposited as droplets and spin-coated onto the substrate surface prior to the CVD process. We formulated several solutions that differ in terms of the inorganic seed promoter in use. In the literature, several studies report the capability of inorganic salts to influence and promote $MoS_2$ growth. In particular, inorganic salts of alkaline metals proved to be utterly interesting and provided remarkable results. One of the most studied salts in literature is sodium chloride (NaCl) which showed the capability of promoting lateral growth, improving the lateral size of the domains and reducing the growth temperature of $MoS_2$ deposition [14]–[16]. In-depth studies suggested that alkaline metals play an effective role in the $MoS_2$ nucleation giving them a potential as outperforming seed promoters [17], [18]. However, no clear mechanism has been proven up to now [19]. In this study, we investigate the role of the following inorganic seed promoters: sodium hydroxide (NaOH), potassium chloride (KCl) potassium iodide (KI) and lithium hydroxide (LiOH).

In the first part of our study, we report on our results with the aim to optimize the growth process in order to grow $MoS_2$ with the lowest possible thickness, which resulted to be two layers (2L). Raman study has been performed on the grown $MoS_2$ to assess the achieved thickness by measuring the frequency difference between the two main vibrational modes ($E^1_{2g}$ in-plane and $A_{1g}$ out-of-plane) of $MoS_2$ and to monitor the effect of the different inorganic seed promoters in terms of uniformity of the grown material and degree of crystalline order. Starting from the Raman characterization of a batch of $MoS_2$ samples grown without the use of a seed promoter, here we provide a comparison with seed promoter-mediated growths supporting the effectiveness of these molecules in granting a higher quality material. Furthermore, in the second part of our study, we investigate the possibility of reducing the deposition temperature while considering the growth of MoS$_2$ in terms of coverage and degree of order. Moreover, a brief analysis on the role of the seed promoter concentration has been conducted.

## 2. MATERIALS AND METHODS

The substrates are cut in 3 cm x 3 cm samples from 4" (100 mm) diameter Si/SiO$_2$ p-doped <100> oriented wafers (Sil'tronix) with 50 nm thick SiO$_2$, cleaned with acetone and isopropanol for, respectively, 180 and 300 seconds and rinsed in deionized (DI) water. Surface treatment with hot Piranha solution (H$_2$O$_2$:H$_2$SO$_4$ = 1:3 v/v) has been performed to hydroxylate the surface and guarantee a successful spin coating of the solution. Three different solutions are prepared containing a) the ammonium heptamolybdate (AHM) Mo precursor in deionized (DI) water, b) an inorganic seed promoter (ISP) in DI water and c) Optiprep®, a density gradient medium based on Iodixanol, which provides viscosity to the solution in order to spin coat it on the substrate surface [20]. The three solutions a),b) and c) are mixed with a volume ratio of 1:1:0.5. The final mixed solutions are five and reported in **Table 1**. The prepared solutions differ for the ISP considered: NaOH, KCl, KI and Li(OH). The quantity of ISP is adjusted to have the same molarity in all the solutions. In one case, for ISP-free MoS$_2$ growths, one solution does not contain any ISP. The so-prepared solutions are spin-coated at 2900 rpm for 60 seconds over the cleaned substrate and then annealed at 120 °C for 60 seconds. Finally, substrates are cleaved to the size of 3 cm x 1 cm. **Figure 1a** shows the schematic view of the single furnace CVD setup used to obtain few-layers MoS$_2$ in a geometry where the spin-coated solution is sulfurized by S powder (250 mg, Acros Organics, 99,999%) evaporation in a process with temperature profile and in N$_2$ flow as shown in the **Figure 1b**. Specifically, the temperature increases linearly at a rate of 5°C/min. In order to achieve a saturated atmosphere in S, at 450°C both the inlet and the outlet are simultaneously closed. The effective growth step is reached at 850°C and lasts 10 minutes. Such high temperature regime guarantees the activation of the MoS$_2$ synthesis. After the growth step, cooling occurs naturally in N$_2$ flux. In-depth analyses of L-CVD grown few-layers MoS$_2$ were performed by micro-Raman spectroscopy (InVia Renishaw, $\lambda_L$ = 514.5 nm, < 10 mW power, 50X, 0.75 N.A.). The MoS$_2$ vibrational modes (E$^1_{2g}$ in-plane and A$_{1g}$ out-of-plane) detected with the Raman spectroscopy are fitted with Voigt functions, a convolution of a Cauchy-Lorentz distribution and a Gaussian distribution, to extrapolate peak position, intensity, full-width-half-maximum (FWHM) and its relative standard error.

| ISP type | Total volume (ml) | AHM (mg) | ISP (mg) | Optiprep (ml) | DI water (ml) | ISP moles (mmol) | ISP molarity (M) |
|---|---|---|---|---|---|---|---|
| NO ISP | 25 | 37,5 | --- | 5 | 20 | --- | --- |
| Na(OH) | 25 | 37,5 | 20 | 5 | 20 | 0,5 | 0,02 |
| KCl | 25 | 37,5 | 37,3 | 5 | 20 | 0,5 | 0,02 |
| KI | 25 | 37,5 | 83 | 5 | 20 | 0,5 | 0,02 |
| Li(OH) | 25 | 37,5 | 12 | 5 | 20 | 0,5 | 0,02 |

**Table** Errore. Nel documento non esiste testo dello stile specificato. – Summary of the solutions formulation. AHM: ammonium heptamolybdate; ISP: inorganic seed promoter; DI: deionized water.

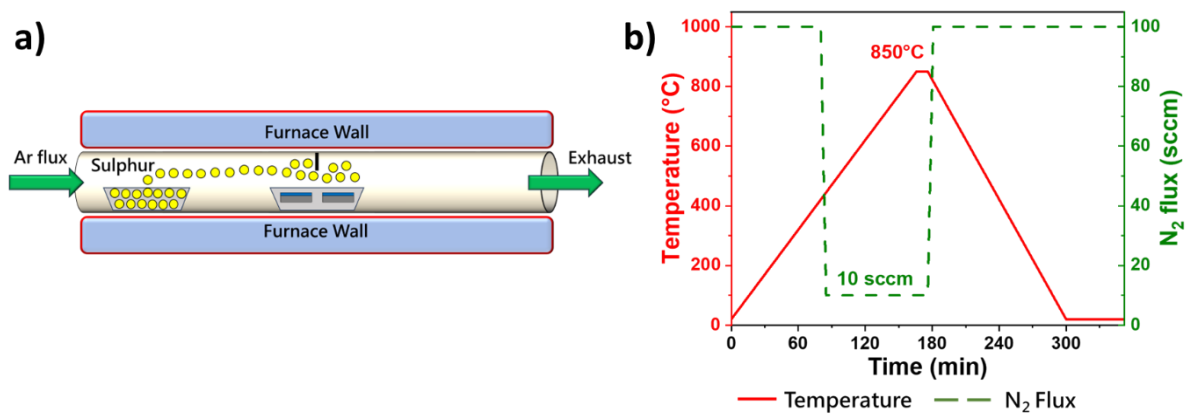

**Figure 1** – a) Single furnace chemical vapour deposition scheme with the exploitation of a liquid precursor synthesis. B) Optimized temperature profile (solid red line) along with N$_2$ flow (dashed green line) versus process time to grow MoS$_2$.

## 3. RESULTS AND DISCUSSION

**Figure 2** shows the intensity of different Raman spectra taken on MoS$_2$ grown with ISP = KCl as a function of the position of the sample where the measurements are taken.

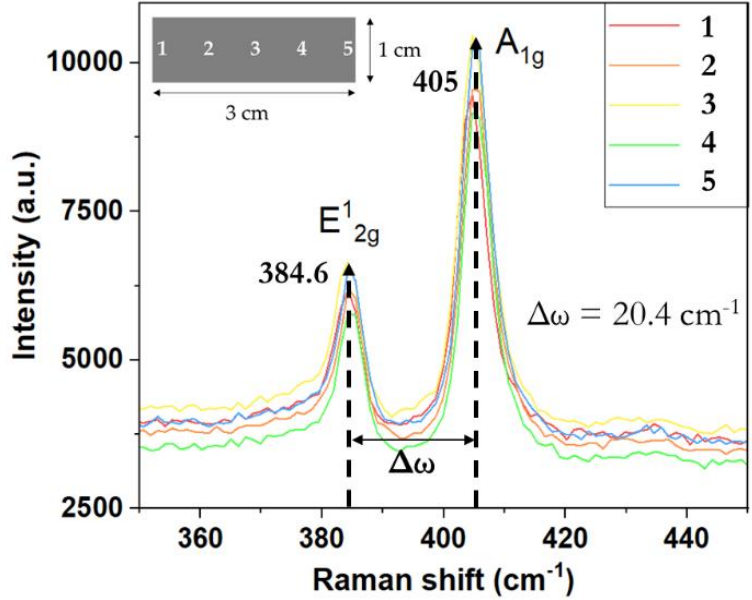

**Figure 2** – Five Raman spectra of KCl-based $MoS_2$ taken at different positions across the surface sample as sketched in the inset on the top-left.

The reported five different measurements have been taken along the 3 cm length of the sample, starting from the sample edge and approximately at a mutual distance of 6 mm, to assess the uniformity of the grown material by considering the frequency shift between $E^1_{2g}$ and $A_{1g}$ mode. The spectra are well overlapped proving that the process grants a very high uniformity of $MoS_2$ growth over $cm^2$ areas, higher when compared to standard CVD processes based on powder precursors. Moreover, the shift between the main phonon vibrational modes $E^1_{2g}$ (in-plane) and $A_{1g}$ (out-of-plane) allows us to evaluate the number of $MoS_2$ layers, as extensively reported in the literature [3], [21], [22]. In this specific case, the measured difference $\Delta\omega$ is equal to 20.4 cm$^{-1}$ which, according to the literature, proves that 2L of $MoS_2$ were grown.

1. *Volume of spin-coated solution*

**Figure 3(a-d)** shows the Raman shift separation between the $E^1_{2g}$ and $A_{1g}$ main phonon modes (cm$^{-1}$), as extracted from the fit to the collected Raman spectra, as a function of the volume of spin-coated

solution in four different cases, depending on the used ISP. Precisely, in **Figure 3**, panel **a)** reports the case when no seed promoter, organic or inorganic, has been used, while the results from ISP-containing solutions are graphed in **b)** for NaOH, **c)** for KCl, **d)** for KI and **e)** for LiOH, respectively. From **Figure 3a**, a linear correlation between the volume of the deposited solution and the grown $MoS_2$ thickness is evident. While depositing 225 µL of solution led to an unsuccessful growth, in the other cases, from a minimum of 3L obtained spin coating 450 µL of solution, a maximum number of layers up to 5 has been reached increasing the volume up to 900 µL. This implies that the molybdenum coming from the AHM precursor and participating in the reaction is the limiting agent, namely the AHM amount determines the condition of S saturation for the reaction with the available Mo. From **Figure 3b**, related to Na(OH)-based $MoS_2$, a minimum thickness of 2 layers was obtained (frequency shift = 20.7 cm$^{-1}$) when 225 µL of Na(OH)-containing solution was deposited on the surface, while the maximum value of 4 layers resulted from changing the volume of spin-coated solution to an amount larger than 450 µL. **Figure 3c** reports an analogous study using KCl as a seed promoter. In this case, 2 layers are grown using 675 µL of solution while lowering or increasing the amount of solution leads to an increase in the frequency distance between $E^1_{2g}$ and $A_{1g}$ modes resulting in an increased thickness. **Figure 3d** reports the data when a KI-based solution is considered; here, a minimum of 2 layers is achieved when 450 and 675 µL of solution are deposited. **Figure 3e** shows that 2L-$MoS_2$ is synthesized with 450 µL and 675 µL of LiOH-based solution. However, while spin-coating 900 µL led to an increased thickness (3L), with a volume equal to 225 µL no successful growth has been achieved. Considering the four plots (**b-e**) related to the use of ISP-based solutions, no clear correlation between the number of grown $MoS_2$ layers and the volume of spin-coated solution is detectable in contrast with the case of solution with only AHM. Moreover, the minimum thickness for Na(OH) is achieved with a much lower volume of solution compared to K-based seed promoters. This discrepancy can be probably ascribed to the usage of ISPs, which actively participate in the reaction mechanism of the $MoS_2$ formation and alter it.

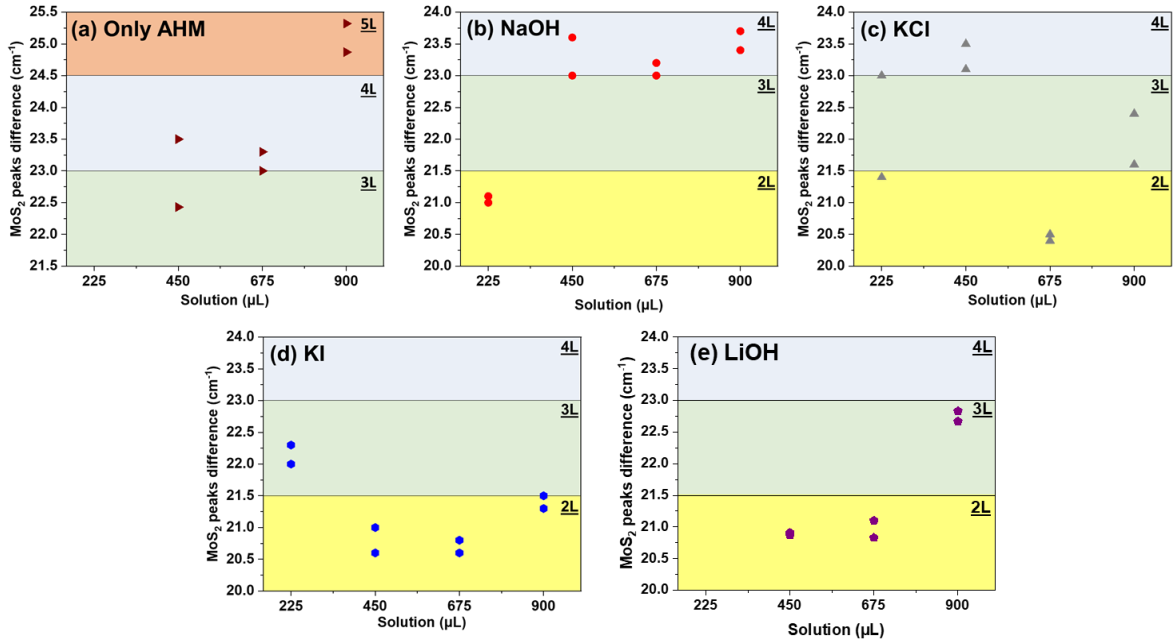

**Figure 3** - (a-e) Correlation graph between the frequency difference of MoS$_2$ main phonon modes from Raman measurements vs. volume of spin-coated AHM (a), AHM + NaOH (b), AHM + KCl (c), AHM + KI (d), and AHM + LiOH (e) solution; measures were taken on 2 samples to confirm the reproducibility.

From the collected data, we can single out an optimal amount of inorganic seed promoter to favour the 2D growth of MoS$_2$ since the decrease or increase in the volume of deposited solution implies that the massive ratio ISP/AHM keeps the same. Moreover, it must be considered that the thickness discrepancies witnessed in samples processed with the same volume of solution probably could be also affected by the spin coating process that could lead to an inhomogeneous distribution over the whole 9 cm$^2$ (3 cm x 3 cm cleaved) substrate surface, therefore resulting in regions of the substrate with a different concentration of Mo precursor and ISP, with a consequence on the non-uniformity in MoS$_2$ growth.

*2. Effect of seed promoter on 3L-MoS$_2$*

**Figure 4** reports the FWHM of E$^1_{2g}$ (a) and A$_{1g}$ (b) MoS$_2$ phonon modes calculated from Raman spectra taken in five positions per sample as reported on the x-axis. To have a consistent and meaningful set of data, all the MoS$_2$ measured samples are characterized by the same thickness equal to 3L. **Table 2** summarizes, for each MoS$_2$ sample, the fitted FWHM from the peaks with the associated relative standard error.

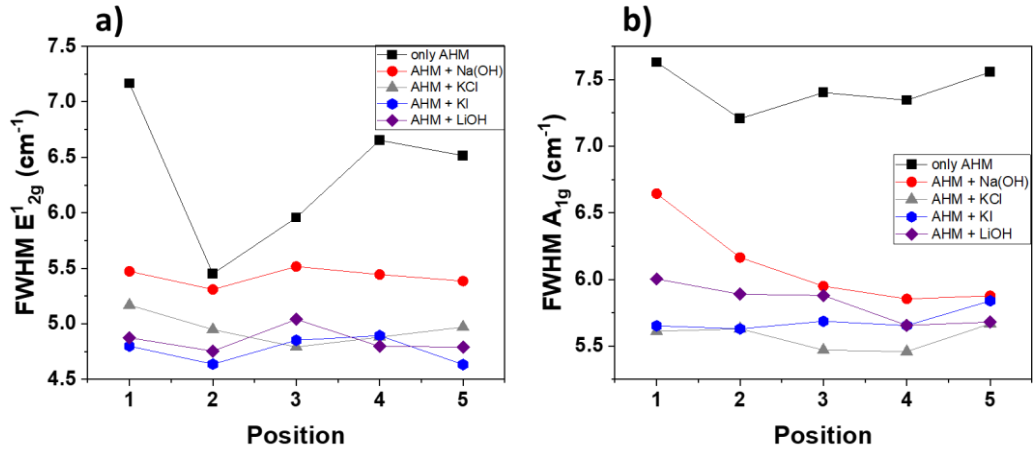

**Figure 4** - a) Full-width half-maximum (FWHM) of E$^1_{2g}$ in-plane MoS$_2$ phonon mode and b) full-width half-maximum of A$_{1g}$ out-of-plane MoS$_2$ phonon mode.

| a) AHM-based 3L-MoS$_2$ | | | | |
|---|---|---|---|---|
| Position | FWHM (E$^1_{2g}$) | Standard error | FWHM (A$_{1g}$) | Standard error |
| 1 | 7,17 | 0,92 | 7,63 | 0,49 |
| 2 | 5,45 | 0,47 | 7,21 | 0,23 |
| 3 | 5,96 | 0,40 | 7,40 | 0,20 |
| 4 | 6,66 | 0,50 | 7,35 | 0,25 |
| 5 | 6,52 | 0,91 | 7,56 | 0,46 |

| b) Na(OH)-based 3L-MoS$_2$ | | | | |
|---|---|---|---|---|
| Position | FWHM (E$^1_{2g}$) | Standard error | FWHM (A$_{1g}$) | Standard error |
| 1 | 5,47 | 0,35 | 6,64 | 0,12 |
| 2 | 5,31 | 0,40 | 6,16 | 0,14 |
| 3 | 5,52 | 0,41 | 5,95 | 0,14 |
| 4 | 5,44 | 0,37 | 5,85 | 0,14 |
| 5 | 5,39 | 0,42 | 5,88 | 0,14 |

| c) KCl-based 3L-MoS$_2$ | | | | |
|---|---|---|---|---|
| Position | FWHM (E$^1_{2g}$) | Standard error | FWHM (A$_{1g}$) | Standard error |
| 1 | 5,17 | 0,43 | 5,61 | 0,15 |
| 2 | 4,95 | 0,46 | 5,63 | 0,16 |
| 3 | 4,79 | 0,43 | 5,47 | 0,16 |
| 4 | 4,88 | 0,48 | 5,46 | 0,17 |
| 5 | 4,97 | 0,45 | 5,67 | 0,16 |

| d) KI-based 3L-MoS$_2$ | | | | |
|---|---|---|---|---|
| Position | FWHM (E$^1_{2g}$) | Standard error | FWHM (A$_{1g}$) | Standard error |
| 1 | 4,80 | 0,52 | 5,65 | 0,18 |
| 2 | 4,64 | 0,49 | 5,63 | 0,17 |
| 3 | 4,85 | 0,50 | 5,69 | 0,18 |
| 4 | 4,89 | 0,51 | 5,65 | 0,18 |
| 5 | 4,63 | 0,48 | 5,84 | 0,17 |

| e) Li(OH)-based 3L-MoS$_2$ | | | | |
|---|---|---|---|---|
| Position | FWHM (E$^1_{2g}$) | Standard error | FWHM (A$_{1g}$) | Standard error |
| 1 | 4,87 | 0,46 | 6,00 | 0,23 |
| 2 | 4,75 | 0,54 | 5,89 | 0,19 |
| 3 | 5,04 | 0,49 | 5,88 | 0,20 |
| 4 | 4,80 | 0,45 | 5,66 | 0,22 |
| 5 | 4,79 | 0,41 | 5,68 | 0,15 |

**Table 2** – FWHM of E$^1_{2g}$ and A$_{1g}$ vibrational modes with respective standard error for 3L-MoS$_2$ obtained from solution with a) no ISP and b) Na(OH), c) KCl, d) KI, e) Li(OH) ISPs.

The black line, with black squared points marking the position of the taken spectra, reports the FWHM of $MoS_2$ grown without ISP use. Considering both the FWHM of $E^1_{2g}$ and $A_{1g}$ phonon modes, the respective values are the highest registered among all $MoS_2$-grown samples. In particular, considering the FWHM of the $E^1_{2g}$ mode, a strong variability is evident going from position 1 (7.2 cm$^{-1}$) and 2 (5.5 cm$^{-1}$) to position 5 (6.5 cm$^{-1}$), suggesting a strong lack of local order in the basal plane of each $MoS_2$ layer. It is worth noticing that this behaviour is not observed in the FWHM of the $A_{1g}$ mode meaning that the layers are well stacked one on top of the other. Moreover, it is possible to appreciate that, considering the data referring to each ISP in solution, the introduction of an inorganic molecule in solution decreases the FWHM values of both $MoS_2$ phonon modes. Considering the FWHM of $E^1_{2g}$ mode for NaOH-based $MoS_2$ (red line with circular dots), an average value of 5.43 cm$^{-1}$ is reported, much lower if compared to the case without the use of a seed promoter, and wider when compared to the values of K-based and Li-based $MoS_2$, the latter almost overlapping as shown in **Figure 4a)**. This difference in FWHM suggests that the degree of local order in the basal plane of each $MoS_2$ layer grown from NaOH is reduced compared with the cases of the same number of $MoS_2$ layers grown from K-based and Li-based salts (KCl in grey with triangular dots, KI in blue with hexagonal dots and Li(OH) in purple with rhomboidal dots). Moving to the FWHM of the $A_{1g}$ mode, a clear distinction between $MoS_2$ grown without and with seed promoters can be appreciated. In addition, the FWHM of NaOH-based 3L-$MoS_2$ spans from 6.7 cm$^{-1}$ in position 1 progressively decreasing to 5.8 cm$^{-1}$ in position 5. In contrast, the FWHM of the $MoS_2$ spectra grown with the other ISP shows an almost constant value despite the measured position. This behaviour suggests a significant lack of uniformity between the $MoS_2$ layers from NaOH-based solutions, where possibly, the average size of the $MoS_2$ domains is progressively increasing from position 1 to position 5. Furthermore, the high degree of uniformity of $MoS_2$ layers from K- and Li(OH)-based ISPs is confirmed.

*3. Effect of lowering the growth temperature*

Having assessed the growth of 2D MoS$_2$ at 850 °C, which we consider as standard deposition temperature (SDT), we investigated the capability of ISPs to permit the controlled growth of MoS$_2$ layers at reduced growth temperatures. From the literature [23] is known that that increasing the temperature up to 270-360°C, depending on the chemical composition of the atmosphere in the furnace, AHM decomposes following the reaction:

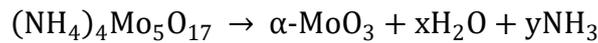

$$(NH_4)_4Mo_5O_{17} \rightarrow \alpha\text{-}MoO_3 + xH_2O + yNH_3$$

where α-MoO$_3$ is the stable structure of molybdenum trioxide. While this temperature range can be taken as a lower limit, the typical reported deposition temperature for MoS$_2$ by CVD ranges between 700 and 800°C with serious effects on the obtained crystalline quality when moving to the lower temperature [24]. Nonetheless, the use of NaCl as ISP has been demonstrated to decrease the growth temperature in 2D MoS$_2$ CVD growth [14]. Taking this into account, we chose to perform different MoS$_2$ growths exploiting KCl with a concentration of 0.05M as a seed promoter at different temperatures to detect the lowest temperature where the growth of MoS$_2$ layers is still granted.

**Figure 5 (a-b)** reports the FWHM of the MoS$_2$ main phonon modes from the growths at 850°C (STD), 750°C, 700°C and 650°C, **Figure 5 (c-d)** reports, respectively, the peak position of the E$^1_{2g}$ and A$_{1g}$ vibrational modes as a function of the growth temperature and the intensities - normalized on Si peak intensity - of both vibrational modes at different growth temperatures. **Figure 5a** shows that the value of the E$^1_{2g}$ peak FWHM reduces at progressively higher temperatures, being 6.4 - 7.5 cm$^{-1}$ for 2L-MoS$_2$ grown at 650°C, 7.0 - 8.7 cm$^{-1}$ at 700°C, around 5.9 cm$^{-1}$ at 750 °C and around 5.0 cm$^{-1}$ at 850°C. This implies that a higher growth temperature improves the degree of local order of the MoS$_2$ domains; furthermore, the reduction of the dispersion value of the FWHM with temperature measured at the different points across the sample surface also implies an enhanced uniform deposition, with implications on the capability to control the MoS$_2$ deposition over large areas. Concomitantly, considering **Figure 5b**, the FWHM of the $A_{1g}$ peak remains almost at the same value despite the process temperature, with values spanning in the 5.5 – 6.5 cm$^{-1}$ range.

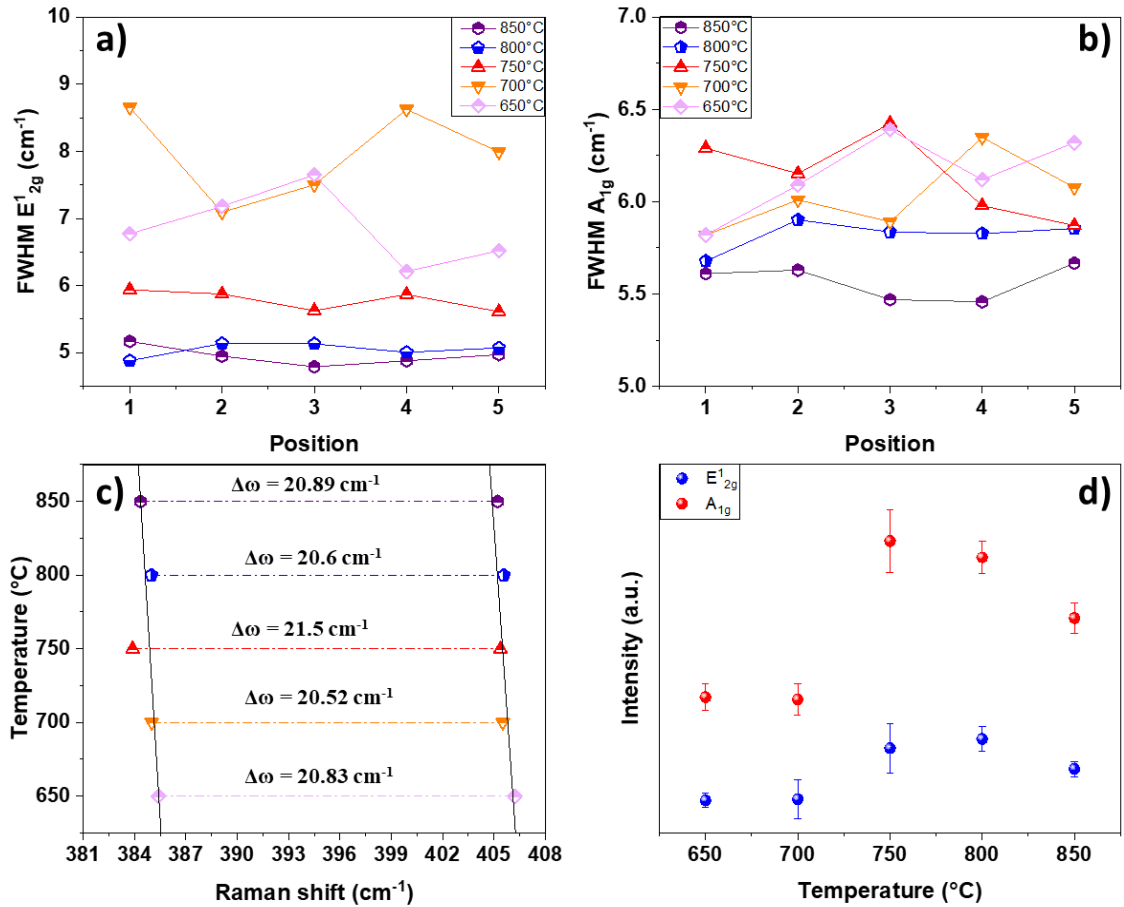

**Figure 5** – (a-b) FWHM of the MoS$_2$ main phonon modes from the growths at 850°C (STD), 800°C, 750°C, 700°C and 650°C, c) $E^1_{2g}$ and $A_{1g}$ peak position at 850°C, 800°C, 750°C, 700°C and 650°C, d) $E^1_{2g}$ and $A_{1g}$ peaks intensity vs deposition temperature. Solid black lines in c) are guides to the eyes.

Such observations could be rationalized as a change in the size and degree of order within each MoS$_2$ layer, while the stacking of the layers one upon the other remains unaffected. In other words, a reduction of the growth temperature limits the size of the MoS$_2$ domains, being few nm$^2$ and disorderly distributed across the surface, and increases the probability of introducing defects at MoS$_2$ domain edges on the surface because of reduced mobility. This justifies the fact that values of FWHM with a minimal dispersion are observed at 850°C claiming for a uniform, regular, MoS$_2$ layer coverage over large areas, which is what we consider as our best-obtained result on the overall quality of MoS$_2$ layers. Considering **Figure 5c**, the position of both $E^1_{2g}$ and $A_{1g}$ peaks shifts to higher frequencies decreasing the growth temperature. In detail, while $E^1_{2g}$ peak position shifts from 384.34 cm$^{-1}$ when MoS$_2$ is processed at 850°C to 385.40 cm$^{-1}$ when MoS$_2$ is synthesized at 650°C, $A_{1g}$ peak position shifts from 405.22 cm$^{-1}$ at 850°C to 406.23 cm$^{-1}$ reducing the temperature at 650°C. In both cases, a redshift is witnessed with an average shift of ~ 1 cm$^{-1}$ which implies that such redshift is not coupled with an increase or decrease of the resulting MoS$_2$ thickness. This rigid shift of the peak positions, shown with guide-to-the-eye lines in **Figure 5c**, could be ascribed to the presence of residual stresses due to the different thermal load that affects the growing MoS$_2$. In fact, the same trend has been observed when ultrathin MoS$_2$ samples are subjected to axial tension that induces a significant tensile strain [25], [26]. Reducing the growth temperature implies that the overall thermal load is consequently reduced and, since cooling occurs in N$_2$ constant flow, also the duration of this step is reduced. These effects could end up in a situation where the relief of the residual thermal stresses is not optimal, which implies that there is a leftover strain of the MoS$_2$ domains in the grown few-layer film. From **Figure 5d,** the intensities of both vibrational modes change with the growth temperature. In particular, a drastic reduction in the intensity of both $E^1_{2g}$ and $A_{1g}$ peaks is witnessed below 750°C. This scenario could be the resulting combination of two mechanisms: firstly, a change in dimensions and distribution of the MoS$_2$ domains due to a lower growth temperature and, secondly, the formation of defects in the layers that act as scattering centres reducing the Raman intensity.

*4. Effect of the seed promoter concentration*

**Figure 6** reports the FWHM of the MoS$_2$ main vibrational modes from the samples synthesized with different KCl concentrations in the solution. FWHM trends in **Figure 6 (a-b)** show a substantial difference in the effect of the seed promoter concentration on the MoS$_2$ vibrational modes. While the $E^1_{2g}$ FWHM is unaffected, within the error, by the change in the concentration of the KCl seed promoter, as shown in **Figure 6a**, the values of $A_{1g}$ vibrational mode, shown in **Figure 6b**, clearly evidence some trends, with the FWHM of the $A_{1g}$ peak increasing non-linearly with the seed promoter concentration. In particular, when the solution is brought from 0.10M to 0.20M, a large increase in the FWHM is measured (from 5.5 cm$^{-1}$ to 6.7 cm$^{-1}$) while between 0.02M and 0.10M, no clear variation is evident. A variability of FWHM of the $A_{1g}$ mode could imply that a distortion of the $MoS_2$ out-of-plane vibration happens when a certain threshold of the seed promoter concentration is exceeded, a possible signature of induced local disorder by the presence of an excess of KCl.

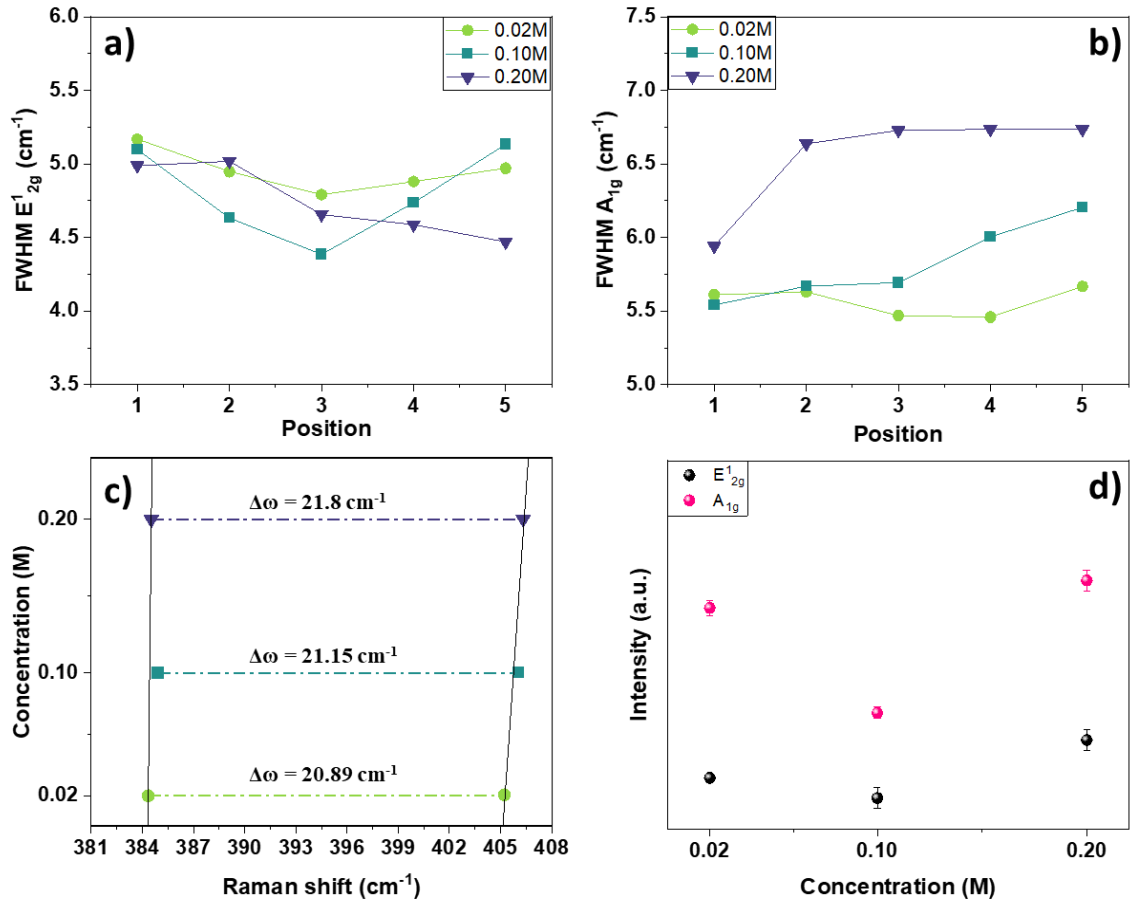

**Figure 6** - 2L-$MoS_2$ processed from solutions with ISP = KCl at different concentrations: a) FWHM of $E^1_{2g}$ mode, b) FWHM of $A_{1g}$ mode, c) $E^1_{2g}$ and $A_{1g}$ peak position in function of the KCl concentration, d) Main vibrational modes intensity at different KCl concentrations. Solid black lines in c) are guides to the eyes.

**Figure 6c** reports the $E^1_{2g}$ and $A_{1g}$ peak positions for few-layers KCl-based $MoS_2$ with increasing seed promoter concentration. As reported in Table 1, the standard KCl solution is the 0.02M one. The position of $E^1_{2g}$ seems not to be affected by the increase of seed promoter concentration while the $A_{1g}$ peak position shows a clear redshifting. Starting from the $A_{1g}$ peak position value equal to 405.22 cm$^-$

$^1$ of 0.02M KCl-based MoS$_2$, for 0.10M KCl-based MoS$_2$ the same peak moved to 406.05 cm$^{-1}$ and for 0.20M the peak position is 406.32 cm$^{-1}$. The total redshift is equal to 1.1 cm$^{-1}$. Rationalizing our result, the so-presented peak trends and the increase in the $\Delta\omega$ are not correlated to the formation of a new layer. A blueshift of the E$^1_{2g}$ peak and a redshift of the A$_{1g}$ peak is reported to occur when a new layer is grown [27]. Therefore, we suppose that the increase of the seed promoter concentration probably enhances the distance between the grown MoS$_2$ layers introducing defectivities that reduce the stacking order of one layer onto the other. In other words, while the order in the basal plane is not affected by the variation of the ISP concentration, the vertical stacking is altered possibly due to the presence of intermediate products or byproducts that lie between one layer and the other, consistently with the increased FWHM shown in **Figure 6 a)** and **b)**.

## 5. CONCLUSIONS

In this study, we report an effective approach based on L-CVD for the synthesis over cm$^2$ lateral areas of 2D-MoS$_2$ nanosheets. With this methodology, it is possible to tailor the number of MoS$_2$ layers by adjusting the volume of the spin-coated solutions. Furthermore, we prove the effectiveness of four different inorganic seed promoters (ISPs) in controlling and promoting the growth of the MoS$_2$ layers and the resulting degree of order at the nanoscale. Among the studied ISPs, K-based salts seem to grant to obtain the highest quality MoS$_2$ growth. In particular, KCl proved to be efficient in reducing the growth temperature down to 650°C with no reduction in the surface coverage and minor effects on the MoS$_2$ quality. We also showed that the concentration of the seed promoter in solution is a key parameter for the controlled synthesis of 2D MoS$_2$ layers since it plays a role in controlling the degree of order and defectivity of the MoS$_2$ layers.

## 6. ACKNOWLEDGEMENTS

The authors acknowledge M. Alia (CNR-IMM) for technical support. Acknowledged financial support for this study comes from *aSTAR* (PRIN grant n. 2017RKWTMY, Ministero dell'Istruzione, dell'Università e della Ricerca, Italy) project.